\documentclass[prb,twocolumn,showpacs,preprintnumbers,amsmath,amssymb]{revtex4}

\usepackage{graphicx}
\usepackage{dcolumn}
\usepackage{bm}
\usepackage{color}

\begin{document}

\preprint{APS/123-QED}

\title{Localization-delocalization transitions in a two-dimensional
quantum percolation model: von Neumann entropy studies}
\author{Longyan Gong  $^{1,}$}
\thanks{Email address:lygong@njupt.edu.cn.}
\author{Peiqing Tong  $^{2,}$ }
\thanks{Corresponding author. Email address:pqtong@njnu.edu.cn.}
\affiliation{
 $^{1}$Center of Optofluidic Technology and College of
 Science, Nanjing University of Posts and Telecommunications, Nanjing 210003,
 China\\
 $^{2}$Department of Physics, Nanjing Normal University, Nanjing 210097,
 China
}%
\date{today}
\begin{abstract}
In two-dimensional quantum site-percolation square lattice models,
the von Neumann entropy is extensively studied numerically. At a
certain eigenenergy, the localization-delocalization transition is
reflected by the derivative of von Neumann entropy which is
maximal at the quantum percolation threshold $p_q$. The phase
diagram of localization-delocalization transitions is deduced in
the extrapolation to infinite system sizes. The non-monotonic
eigenenergies dependence of $p_q$ and the lowest value
$p_q\simeq0.665$ are found. At localized-delocalized transition
points, the finite scaling analysis for the von Neumann entropy is
performed and it is found the critical exponents $\nu$ not to be
universal. These studies provide a new evidence that the existence
of a quantum percolation threshold $p_q<1$ in the two-dimensional
quantum percolation problem.
\end{abstract}
\pacs{71.30.+h, 03.67.-a, 72.15.Rn}%
\maketitle

\section{Introduction}
The Anderson model\cite{an58,ev08} and the quantum percolation
(QP) model\cite{ki72,de83} are two important theoretical models
that are used to study electron localization properties in
disordered systems. In the two models, the on-site energy
randomness and the geometric randomness are considered,
respectively. For the Anderson model, there are extensive studies
and definitive results \cite{ev08}, while for the QP model, there
are many open issues even today.

 The main concern in QP problems is to locate the QP
threshold $p_q$ (accessible site concentrations by quantum
particles) below which the electron is localized with probability
one. For the Anderson model and the QP model, there is a consensus
on the existence of localization-delocalization transitions (LDTs)
in three dimensions (3D) \cite{so87,ko91,be96,sc05}. For Anderson
models, according to the one-parameter scaling theory\cite{ab79},
LDTs do not occur at and below two dimensions (2D). However,
whether the scaling theory is suitable for the QP model is an open
question \cite{sc08} and even whether LDTs exist in 2D QP models
at $p_q<1$ is less clear \cite{is08}. Studies such as the scaling
work based on numerical calculations of the conductance
\cite{av92} and transfer matrix methods with finite-size scaling
analysis \cite{so91} showed no evidence for LDTs. At the same
time, there are many studies that claim LDTs
exist\cite{od80,ko90,sr84,od84,ra84}. However, the values of QP
threshold $p_q$ obtained by different methods are not consistent.
For example, for 2D quantum site-percolation square lattice
models, Odagaki, \emph{et al.}, obtained $p_q\simeq0.59$  by a
Green's function method\cite{od80}. Koslowski and von Niessen gave
$p_q\simeq0.70$ with the Thouless number based on the
Thouless-Edwards-Licciardello method \cite{ko90}. Srivastava and
Chaturvedi showed $p_q\simeq0.73$ with the participation used by
the method with equations of motion \cite{sr84}. Odagaki and Chang
found $p_q\simeq0.87$ using a real space renormalization-group
method \cite{od84}. Raghavan obtained $p_q\simeq0.95$ by mapping a
2D system into a one-dimensional system \cite{ra84}. Very
recently, Islam and Nakanishi suggested that $p_q$ depending on
particle energies by calculating the transmission coefficient for
2D bond-percolation models\cite{is08}.

In the mean time, metal states or metal-insulator transitions are
observed experimently in a variety of dilute two-dimensional
electron and hole systems\cite{ab01}.  At the same time, the
unusual transport properties of novel materials \cite{sc08}, such
as metal-insulator transitions happen in perovskite manganite
films \cite{zh02} and in granular metals \cite{fe04}, and minimal
conductivity in undoped graphenes \cite{ch07}, may be explained by
2D QP models. Therefore these give additional motivations to study
LDTs in 2D QP models.

On the other hand,  quantum entanglement, which attracting much
attention in quantum information\cite{bo00}, has been extensively
applied in condensed matter
physics\cite{be96,za02,gu04,mu09,ch06,go06,go07,go08,ji08,os02,su06}.
For example, quantum entanglement measured by the von Neumann
entropy was studied in the extended Hubbard model\cite{gu04,mu09},
in quantum small-world networks\cite{go06}, in two interacting
particle systems\cite{go07}, in the extended Harper
model\cite{go08,mu09}, in three dimensional Anderson
models\cite{ji08}, in the integer quantum Hall system
\cite{ji08}, and in spin models \cite{os02,su06}. 
It is found that the von Neumann entropy shows singular behaviors
at quantum critical points(QCPs). The derivative of von Neumann
entropy has very good finite size scaling behaviours  close to
QCPs even for quite small system sizes \cite{ji08,os02,su06}.
Therefore it becomes a powerful method to quantify QCPs in various
systems.


In this paper, with the help of the von Neumann entropy we present
a detailed numerical study of LDTs in the 2D quantum
site-percolation model . Our studies show that a quantum
site-percolation threshold $p_q<1$ exists in the 2D QP problem. In
the next section the QP model and the definition of von Neumann
entropy are introduced. In Sec.~\ref{sec3} the numerical results
are presented. And we present our conclusions and discussions in
Section ~\ref{sec4}.

\section{\label{sec2}Quantum site-percolation model and von Neumann entropy}

\subsection{\label{sec21} Quantum site-percolation model}
Let us consider one-electron tight-binding Hamiltonian with
diagonal disorder defined on square lattices of sites \cite{sc08}
 The on-site potential $\xi_i$ can be drawn from the bimodal
distribution
\begin{equation}
 p(\xi_i)=p\delta(\xi_i-\xi_A)+
 (1-p)\delta(\xi_i-\xi_B).\label{form2}
\end{equation} 
In the limit $\xi_B-\xi_A \longrightarrow\infty$, the electron
moves only on a random assembly of A-lattice points. Without loss
of generality we choose $\xi_A=0$. In the situation the A-site
occupation probability is $p$ and the corresponding quantum
site-percolation Hamiltonian reads
\begin{equation}
H_{AA}=-t\sum\limits_{ \langle{ij} \rangle\in
A}(c^{\dag}_{i}c_{j}+H.c.),\label{form3}
\end{equation} 
where the summation extends over nearest-neighbor A-sites only.

\subsection{\label{sec24} von Neumann entropy}
The general definition of entanglement is based on the von Neumann
entropy \cite{be96von}. The generic eigenstate
$\left|\alpha\right\rangle$ for Hamiltonian (\ref{form3}) with
eigenenergy $\varepsilon_{\alpha}$ is the superposition
\begin{equation}
\left|\alpha\right\rangle=\sum\limits_{i\in A}\psi^{\alpha}_i
\left|i\right\rangle=\sum\limits_{i\in A}\psi^{\alpha}_i
c^{\dag}_{i} \left|0\right\rangle,\label{form4}
\end{equation} 
where $\left|0\right\rangle$ is the vacuum and $\psi^{\alpha}_i$
is the amplitude of wave function at \emph{i}th site. For an
electron in the system, there are two local states at each site,
i.e., $\left| 1 \right\rangle_i$ and $\left| 0 \right\rangle_i$,
corresponding to the state with (without) an electron at the
\emph{i}th site, respectively. The local density matrix $\rho_i$
is defined \cite{za02,gu04,go06,go07,go08,ji08} by
\begin{equation}
\rho_i= z_i\left| {1} \right\rangle{_i}{_i}\left\langle {1}
\right| + (1-z_i)\left| {0} \right\rangle{_i}{_i}\left\langle {0}
\right|,\label{form5}
\end{equation} 
where $z_i=\left\langle \alpha  \right|c_i^ \dag  c_i \left|
\alpha \right\rangle=\left|\psi^\alpha _i \right|^2$ is the local
occupation number at \emph{i}th site. Consequently, the
corresponding von Neumann entropy related to the \emph{i}th site
is
\begin{equation}
E^\alpha_{vi}=-z_i\log_2z_i-(1-z_i)\log_2(1-z_i).\label{form6}
\end{equation}
For nonuniform systems, the value of $E^\alpha_{vi}$ depends on
the site position $i$. At an eigenstate $\alpha$, we define a
site-averaged von Neumann entropy
\begin{equation}
E^\alpha_v= \frac{1}{N} \sum\limits_{i=1}^N
{E^\alpha_{vi}},\label{form7}
\end{equation} 
were $N$ is the number of A-lattice sites. The definition
(\ref{form7}) shows that for an extended state that
$\psi^\alpha_{i}=\frac{1}{\sqrt{N}}$ for all $i$,
 $E^\alpha_v=-\frac{1}{N}\log_2 \frac{1}{N}- (1-\frac{1}{N})\log_2
(1-\frac{1}{N}) \approx \frac{1}{N}\log_2{N}$ at $N
\longrightarrow\infty$, and for a localized state that
$\psi^\alpha_i=\delta_{ii_0}$( $i_0$ is a given site ) ,
$E^\alpha_v=0$. In the present paper all the values of
$E^\alpha_v$ is scaled by $\frac{1}{N}\log_2{N}$. From the two
examples, we know the scaled $E^\alpha_v$ is near $1$ when
eigenstates are extended, and near zero when eigenstates are
localized. Henceforth, we omit ``scaled'' for simplicity.  

For a random system the site-averaged von Neumann entropy
$E^\alpha_v$ should be further averaged over different
realizations of disorder. The resulting quantity, the disorder
averaged von Neumann entropy denoted by $\langle
E^\alpha_v\rangle$ for eigenenergy $\varepsilon_\alpha$, which is
defined as
\begin{equation}
 {\langle E^\alpha_v\rangle}=\overline{\frac{1}{N} \sum\limits_{i=1}^N
{E^\alpha_{vi}}}=\frac{1}{K}\frac{1}{N}\sum\limits_{i=1}^N
{E^\alpha_{vi}},\label{form8}
\end{equation} 
where $\overline{X}$ is denoted as random average, $K$ is the
number of disorder realizations. In practice, $\langle
E^\alpha_v\rangle$ is the average values over a small window
$\Delta$ around an energy value $\varepsilon$, i.e.,
$\varepsilon_\alpha\in[\varepsilon-\Delta/2,\varepsilon+\Delta/2]$.
We ensure that $\Delta$  is sufficiently small, and at the same
time there are enough states in the interval $\Delta$. Here
$\Delta=0.04$ is chosen and other $\Delta$ give similar results.

Another quantitative measure that is widely used to characterize
localization is the participation ratio(PR)\cite{be70}, defined by
$\xi_\alpha=(N\sum\limits_{i=1}^N|\psi^{\alpha}_i|^4)^{-1}$, which
gives the ratio of lattice sites occupied by particles to all
lattice sites at an eigenstate $\alpha$. For the above extended
and fully localized states, $\xi_\alpha=1$ and $1/N$,
respectively. Generally speaking, the larger $\xi_\alpha$ is, the
more delocalized the eigenstate is. Similarly, the  disorder
averaged PR $\langle \xi_\alpha\rangle$ is also averaged over the
same small energy window. Henceforth, the disorder averaged von
Neumann entropy $\langle E^\alpha_v\rangle$ and the disorder
averaged PR $\langle \xi_\alpha\rangle$ are simplified to the von
Neumann entropy $\langle E^\alpha_v\rangle$ and the PR $\langle
\xi_\alpha\rangle$, respectively.

\section{\label{sec3} numerical results}
In numerical calculations, we directly diagonalize the eigenvalue
Eq.(\ref{form3}) with the periodic boundary condition and obtain
all eigenvalues $\varepsilon_\alpha$ and the corresponding
eigenstates $\left|\alpha\right\rangle$. Without loss of
generality, the hopping integral $t$ is taken as units of energy.
From formulas (\ref{form6}-\ref{form8}), we can obtain the von
Neumann entropy $\langle E^\alpha_v\rangle$.  We consider systems
of linear size $L=20,30,...,60$ (measured in units of the lattice
constant) and having $N = L^2$ A-lattice sites in total. The
corresponding number of disorder realizations
$K=2500,2000,...,500$, respectively. More realizations give
similar results.

\begin{figure}[!htbp]
(a)\includegraphics[width=2.5in]{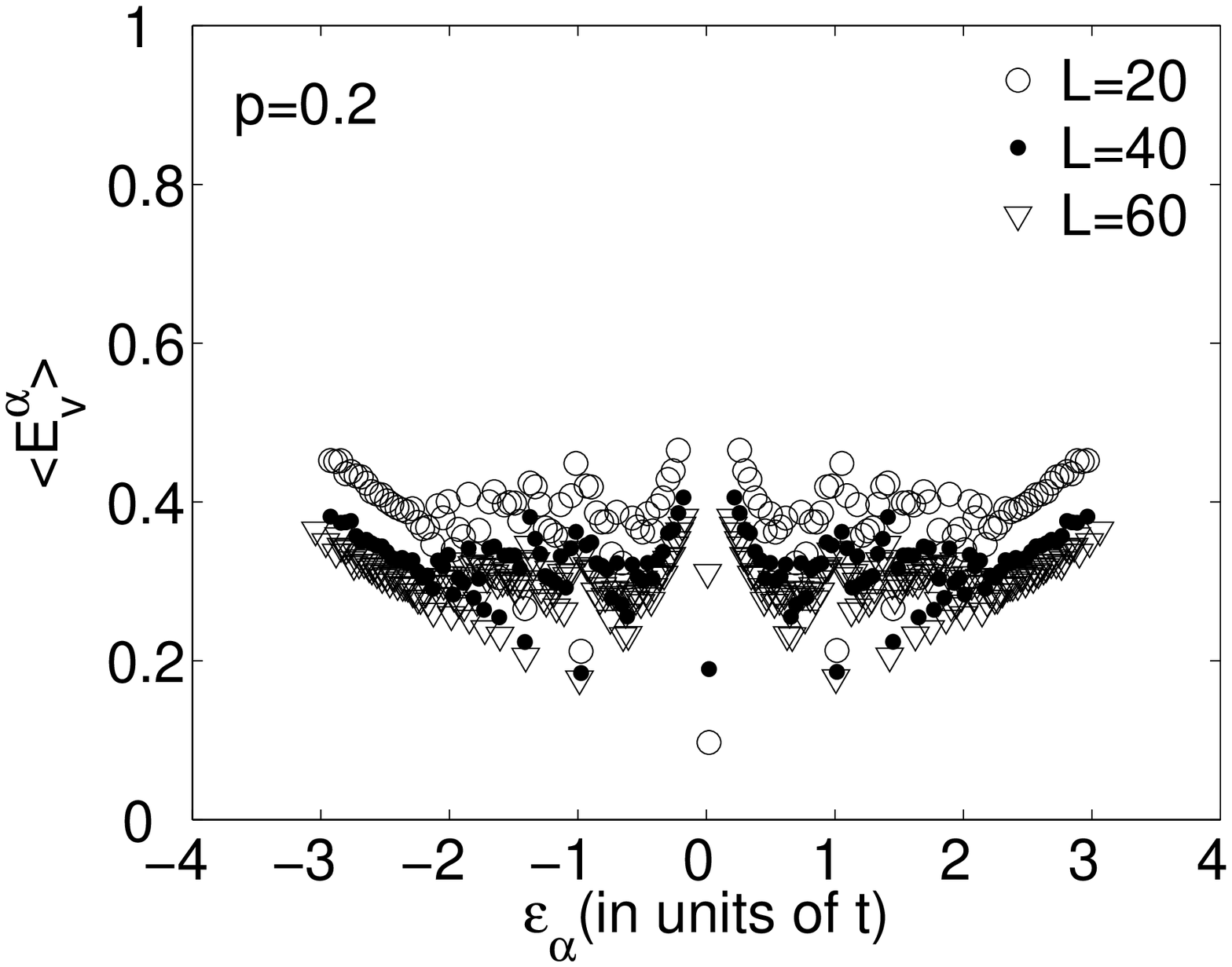}
(b)\includegraphics[width=2.5in]{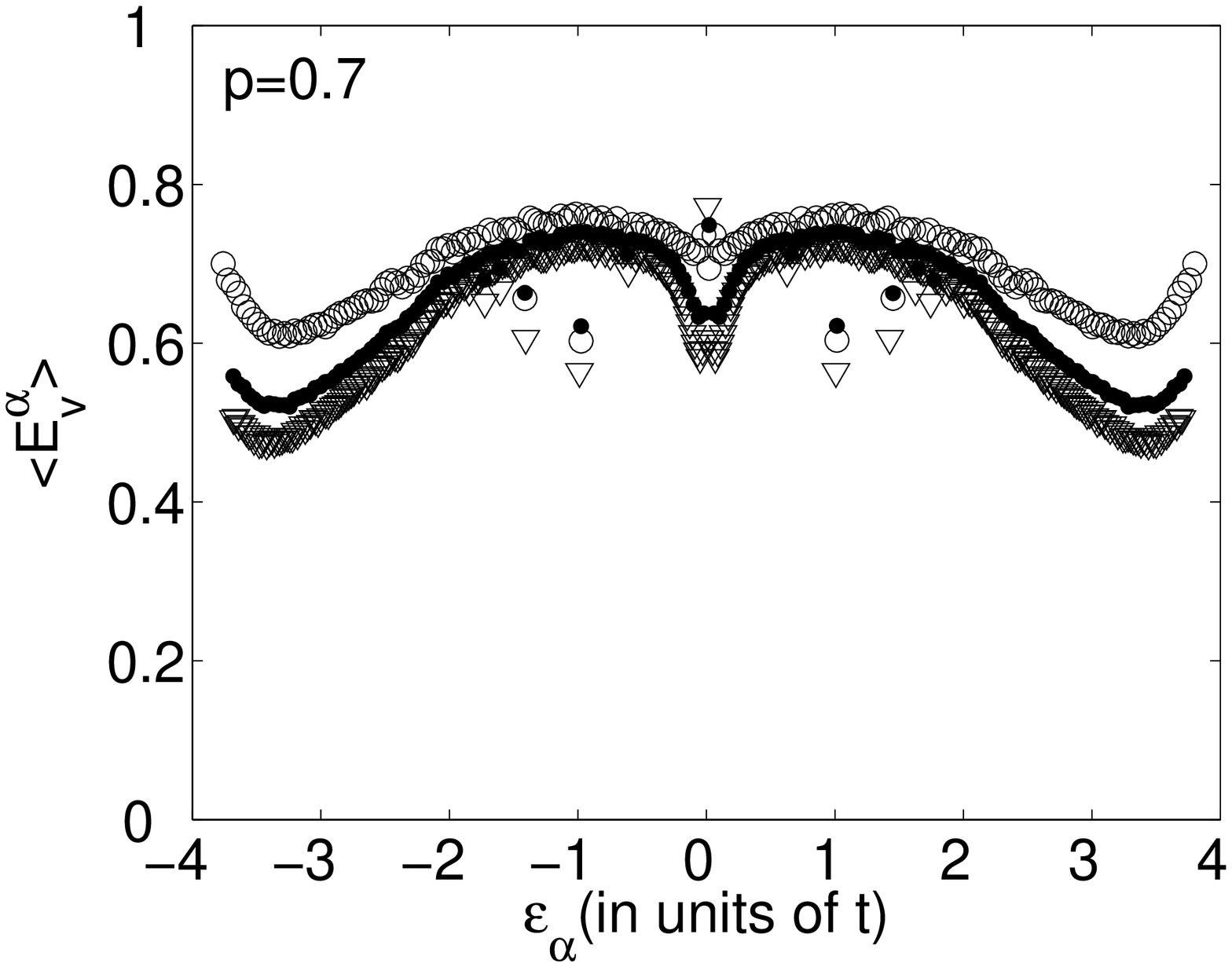}
(c)\includegraphics[width=2.5in]{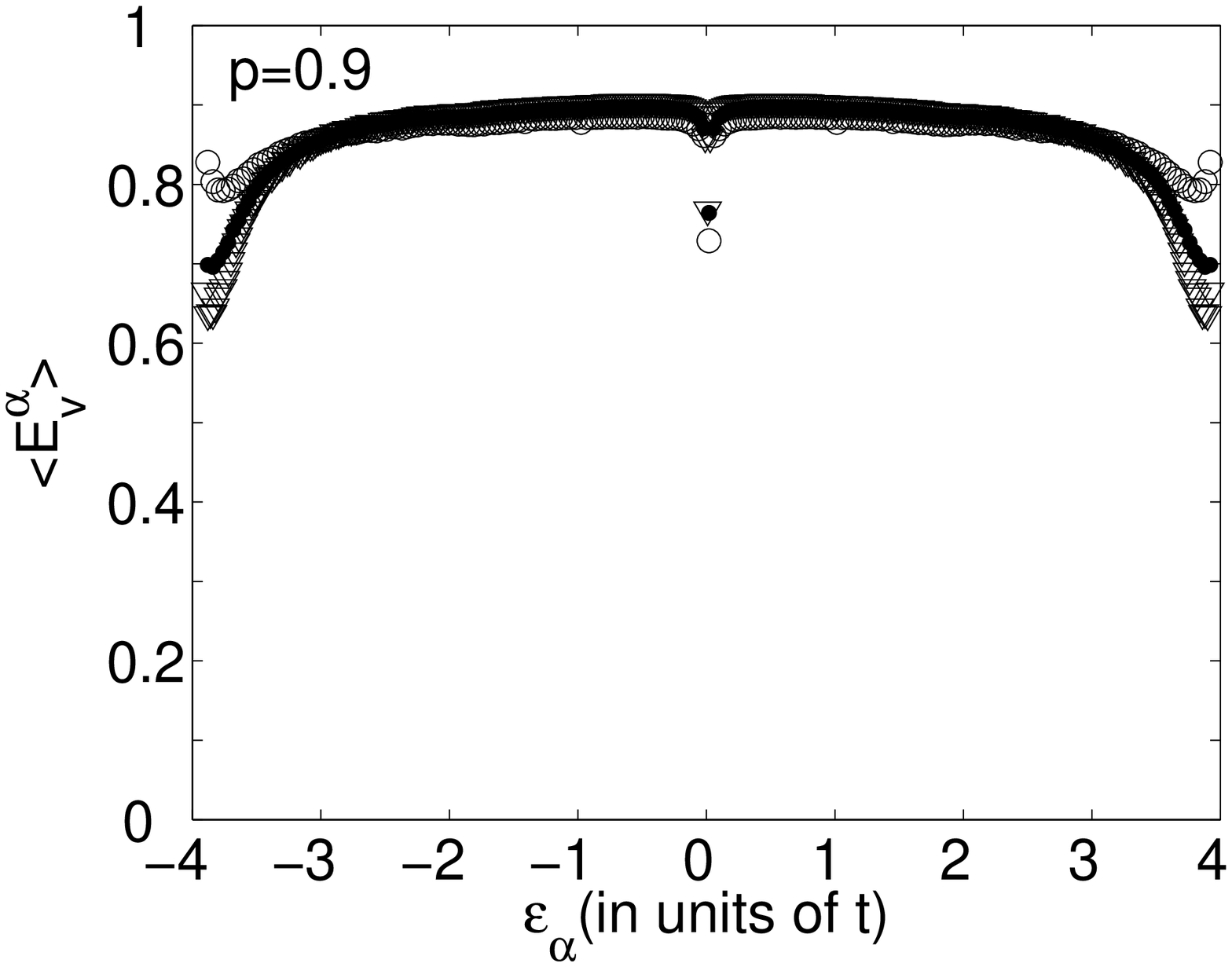}
(d)\includegraphics[width=2.5in]{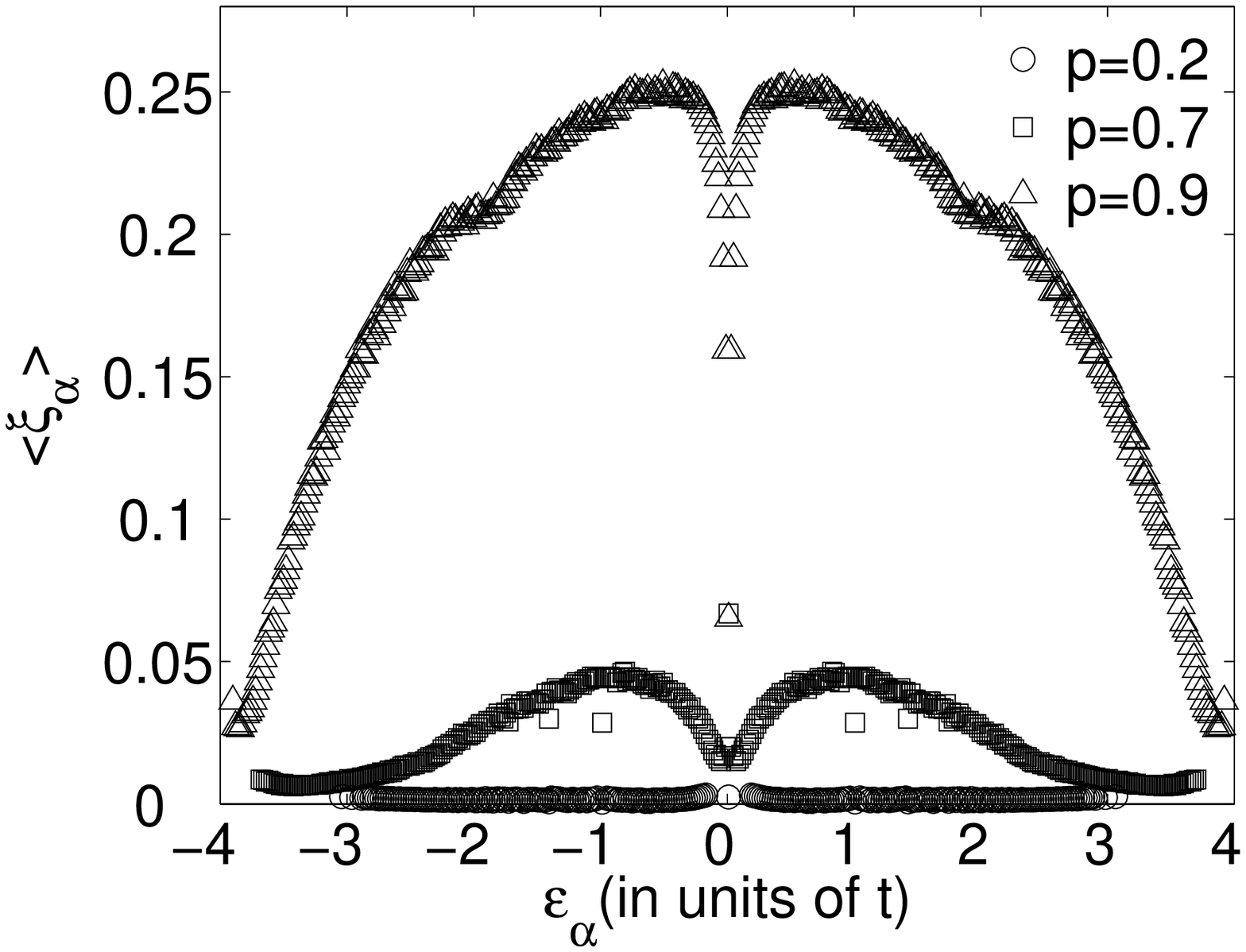}
 \caption{The von Neumann entropy $\langle E^\alpha_v\rangle$  varying with
eigenenergies $\varepsilon_\alpha$ at $L=20,40$ and $60$ for  site
occupation probability (a)$p=0.2$,(b)$p=0.7$ and (c)$p=0.9$,
respectively. (d)The corresponding PR $\langle \xi_\alpha\rangle$
at $p=0.2,0.7$, and $0.9$ for $L=60$.}\label{Fig1}
\end{figure}

\begin{figure}
 \includegraphics[width=2.5in]{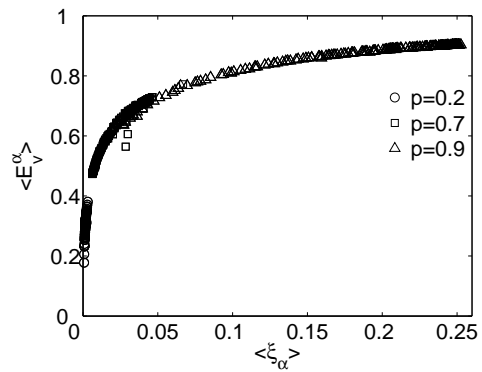}
 \caption{The relation between the $\langle E^\alpha_v\rangle$ and the
corresponding PR $\langle \xi_\alpha\rangle$ at $L=60$ for states
shown in Fig.\ref{Fig1}. }\label{Fig2}
\end{figure}

Before discussing possible localized-delocalized transitions let
us investigate the behavior of the von Neumann entropy $\langle
E^\alpha_v\rangle$ at different site occupation probability $p$.
The $\langle E^\alpha_v\rangle$ as functions of eigenenergies
$\varepsilon_\alpha$ are plotted in Figs.\ref{Fig1}(a)-(c) at
$L=20,40$ and $60$ for $p=0.2,0.7$ and $0.9$, respectively. It
shows that at the same $p$, the values of $\langle
E^\alpha_v\rangle$ depend on eigenenergies $\varepsilon_\alpha$
and system sizes $L$. For $p=0.7$ and $0.9$, the $\langle
E^\alpha_v\rangle$ are relatively small for eigenstates near the
band edges,  while relatively large for eigenstates near the band
center except eigenstates very near the band edges. Comparing the
values of $\langle E^\alpha_v\rangle$ for the three $p$ at the
same $L$, on the whole, all $\langle E^\alpha_v\rangle$ are
relatively small for $p=0.2$ and relatively large for $p=0.9$.
Compared to our results, the PR are also plotted in
Fig.\ref{Fig1}(d) for $L=60$ at the three $p$. The variations of
the PR $\langle\xi_\alpha\rangle$ with respect to
$\varepsilon_\alpha$ are similar to that of $\langle
E^\alpha_v\rangle$ in Figs.\ref{Fig1}(a)-(c). At the same time,
the $\langle E^\alpha_v\rangle$ versus the corresponding
$\xi_\alpha$ are plotted in Fig.\ref{Fig2}. It shows that the
$\langle E^\alpha_v\rangle$ increases monotonously with the
$\langle\xi_\alpha\rangle$, so the von Neumann entropy can reflect
the localization properties of eigenstates in the QP model.

To exact locate localized-delocalized transition points and the
corresponding critical exponents with a finite-size scaling
analysis, in the follows, we study the von Neumann entropy
$\langle E^\alpha_v\rangle$ changes with the site occupation
probability $p$ for different eigenenergies . We found that the
general trend for all von Neumann entropy $\langle
E^\alpha_v\rangle$ curves at different eigenenergies are similar
except the one at the band center , so we will discuss only two
energies as examples, one that is away from the band center and
one very close to the band center.

\subsection{\label{sec31}Eigenenergies away from the band center}

We first study the von Neumann entropy $\langle E^\alpha_v\rangle$
and related quantities for eigenstates with eigenenergies
 $\varepsilon_\alpha\in[-0.84-0.02,-0.84+0.02]$. The
 literatures \cite{is08,od80,ko90,sr84,od84} agree on the QP
threshold $p_q\geq p_c\simeq0.593$ if $p_q$ exits, where $p_c$ is
the classical percolation threshold \cite{st92}. Therefore, we
study the QP model at the site occupation probability $p$
beginning from $0.4$, which is far smaller than the lower bound of
$p_q$.

\begin{figure}
(a)\includegraphics[width=2.5in]{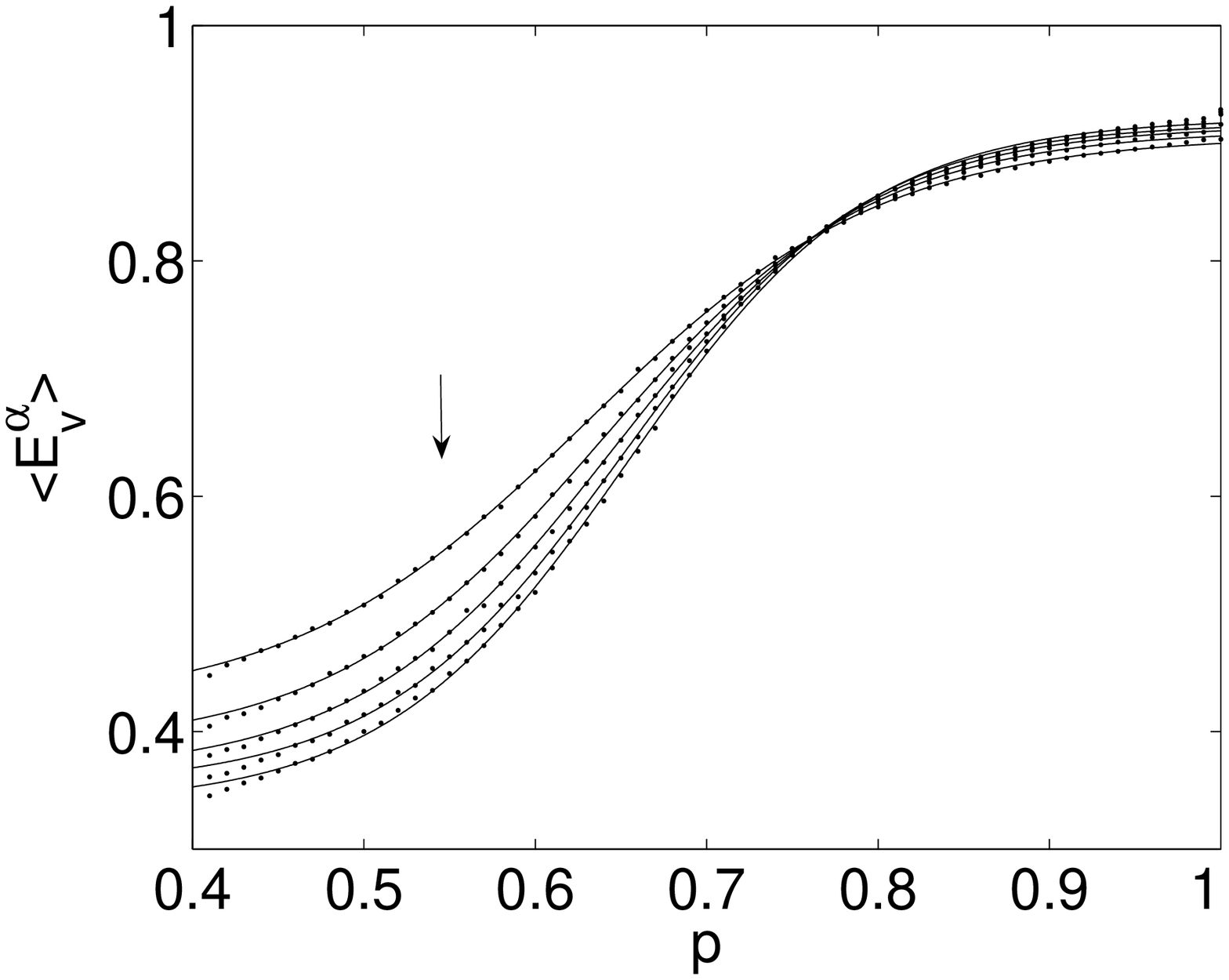}
(b)\includegraphics[width=2.5in]{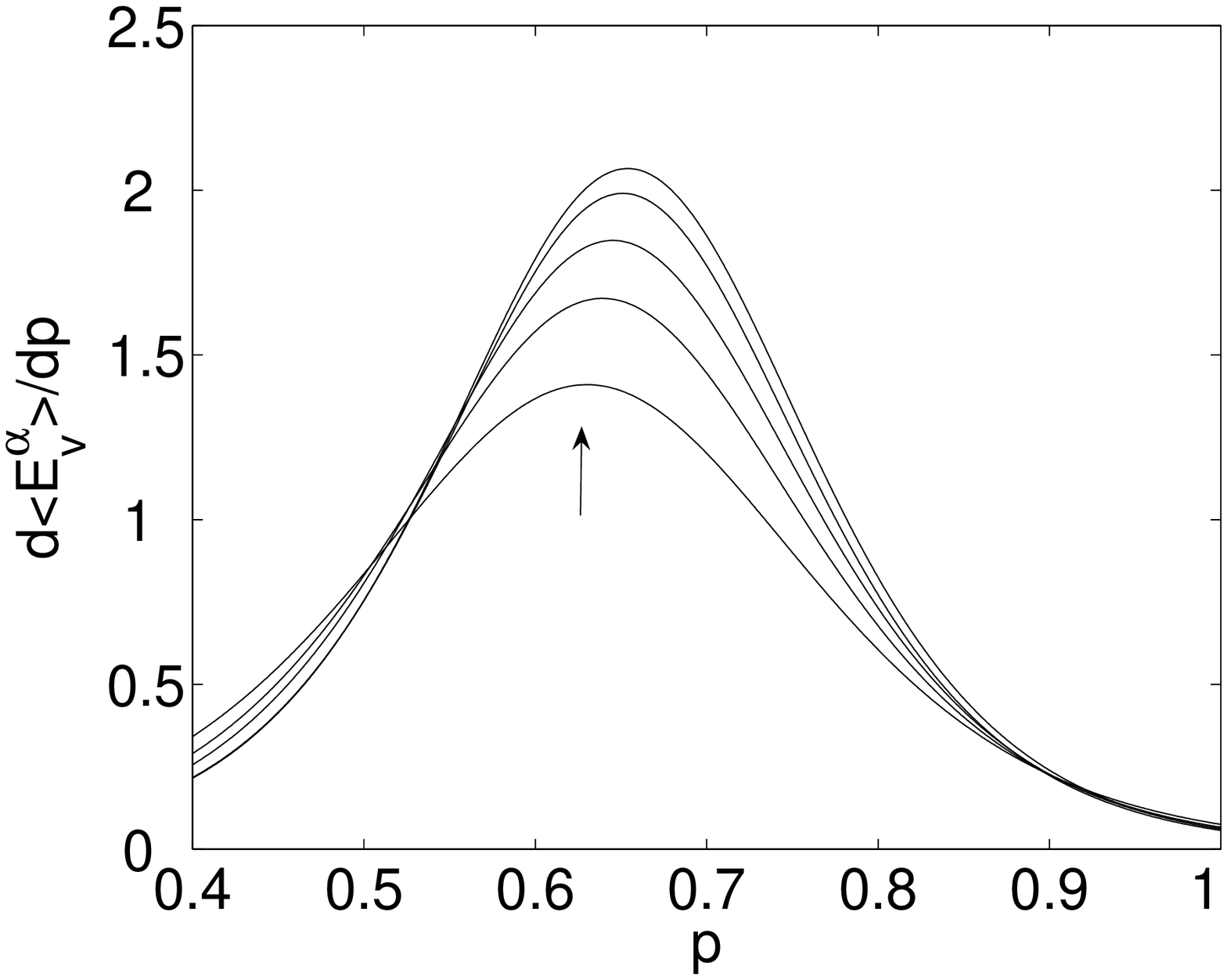}
(c)\includegraphics[width=2.5in]{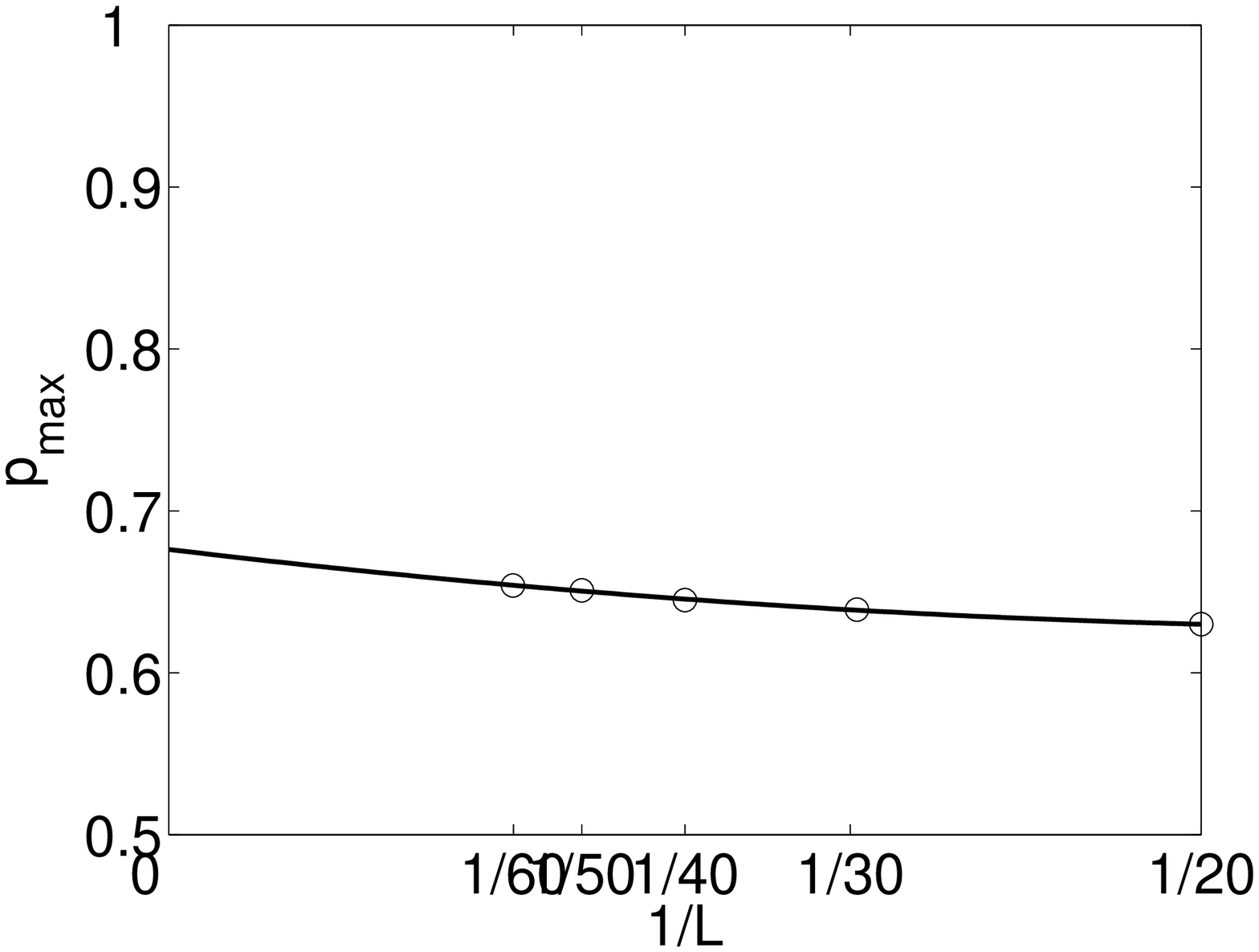}
(d)\includegraphics[width=2.5in]{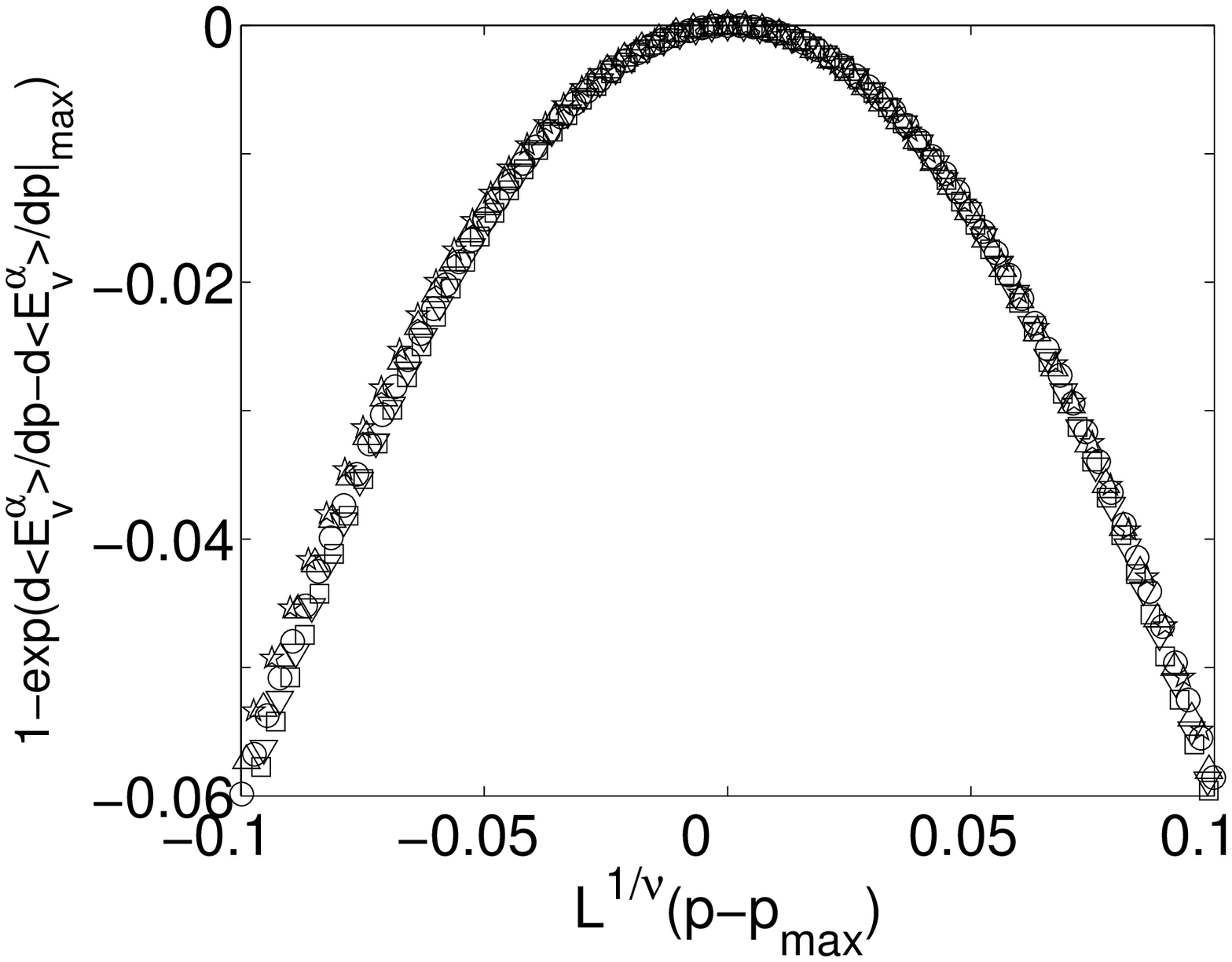}
 \caption{Some quantities for eigenstates with eigenenergies
 $\varepsilon_\alpha\in[-0.84-0.02,-0.84+0.02]$. (a) The
von Neumann entropy $\langle E^\alpha_v\rangle$  varying with  the
site occupation probability $p$ and the lines are Boltzmann
fitting. (b) The derivative $d\langle E^\alpha_v \rangle/dp$
varying with $p$. (c) The line corresponds to the expected
behavior of $p_{max}$ for $1/L\longrightarrow0$ according to a
second-order polynomial fitting. (d) The finite-size scaling
analysis. The system sizes $L=20,30,...,60$ and the arrow
direction in (a) and (b) denotes the increasing of $L$.
}\label{Fig5}
\end{figure}

In Fig.\ref{Fig5}(a), we show the dependence of $\langle
E^\alpha_v\rangle$ on the site occupation probability $p$ at
system sizes $L=20,30,...,60$, respectively.  It shows that
$\langle E^\alpha_v\rangle$ monotonically increases as $p$ becomes
larger. For a certain system size, when $p$ is small, e.g.,
$p=0.4$, the eigenstates are localized and $\langle
E^\alpha_v\rangle$ is small. When $p=1.0$, the model shown in
Eq.(\ref{form3}) is a two-dimensional periodic potential system.
Due to the Bloch theorem the eigenstate of a tight-binding
electron on a local regular lattice is always in the extended
state. At the situation, $\langle E^\alpha_v\rangle$ is largest.
All these reflect the trivial delocalization effect of $p$, which
is similar as that studied in quantum small-world network
models\cite{go06}. All data shown in Fig.\ref{Fig5}(a) are well
fitted with nonlinear Boltzmann functions for various system
sizes. According to the fitted lines, we plot the derivative
$d\langle E^\alpha_v \rangle/dp$ varying with $p$ in
Fig.\ref{Fig5}(b). It shows there is a peak in the derivative at a
certain $p$, which is denoted by $p_{max}$. The maximal derivative
and $p_{max}$ increase with the system sizes $L$, respectively. It
is believed that the von Neumann entropy may be non-analytic at a
quantum phase transitions and can reflect various quantum critical
points\cite{gu04,ji08}. Therefore, that the derivative is maximal
at a certain position $p_{max}$, can be as a signature of LDTs of
electron states\cite{mu09,go06}.

To study the LDTs at the QP threshold $p_q$, one needs to
investigate the behavior of systems in the thermodynamic limit.
However, in most cases this is not possible in numerical
methods\cite{co01}, and therefore, similarly as in
Ref.\onlinecite{mu09} an extrapolation method is chosen.
Fig.\ref{Fig5}(c) shows the scaling behavior of the $p_{max}$. The
$p_q$ in the thermodynamic limit can be obtained by
$1/L\longrightarrow0$ and Fig.\ref{Fig5}(c) shows $p_q\simeq
0.676$ at the situation.  We denote the derivative of von Neumann
entropy at $p_{max}$ as $d\langle E^\alpha_v\rangle/dp|_{max}$ .
Following the Refs. \onlinecite{os02} and \onlinecite{su06},  the
finite size scaling is performed for the function $1-exp(d\langle
E^\alpha_v\rangle/dp-d\langle E^\alpha_v\rangle/dp|_{max})$ with
respect to $L^{1/\nu}(p-p_{max})$. The result is presented in
Fig.\ref{Fig5}(d). It shows numerical results obtained from
various system sizes approximately collapse on a single curve with
the critical exponent $\nu\simeq 2.52$.  

\subsection{\label{sec32} Eigenenergies near the band center}

\begin{figure}
(a)\includegraphics[width=2.5in]{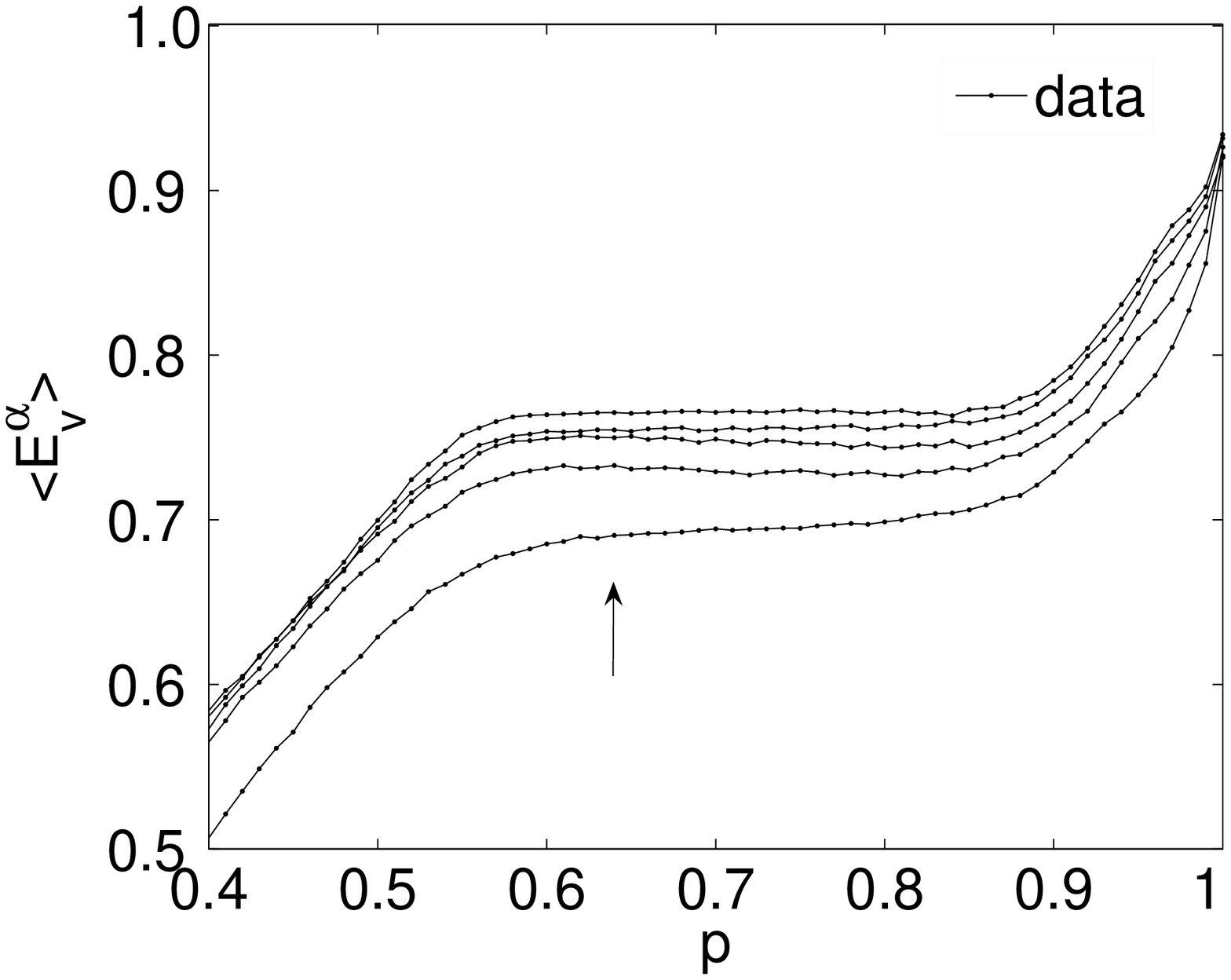}
(b)\includegraphics[width=2.5in]{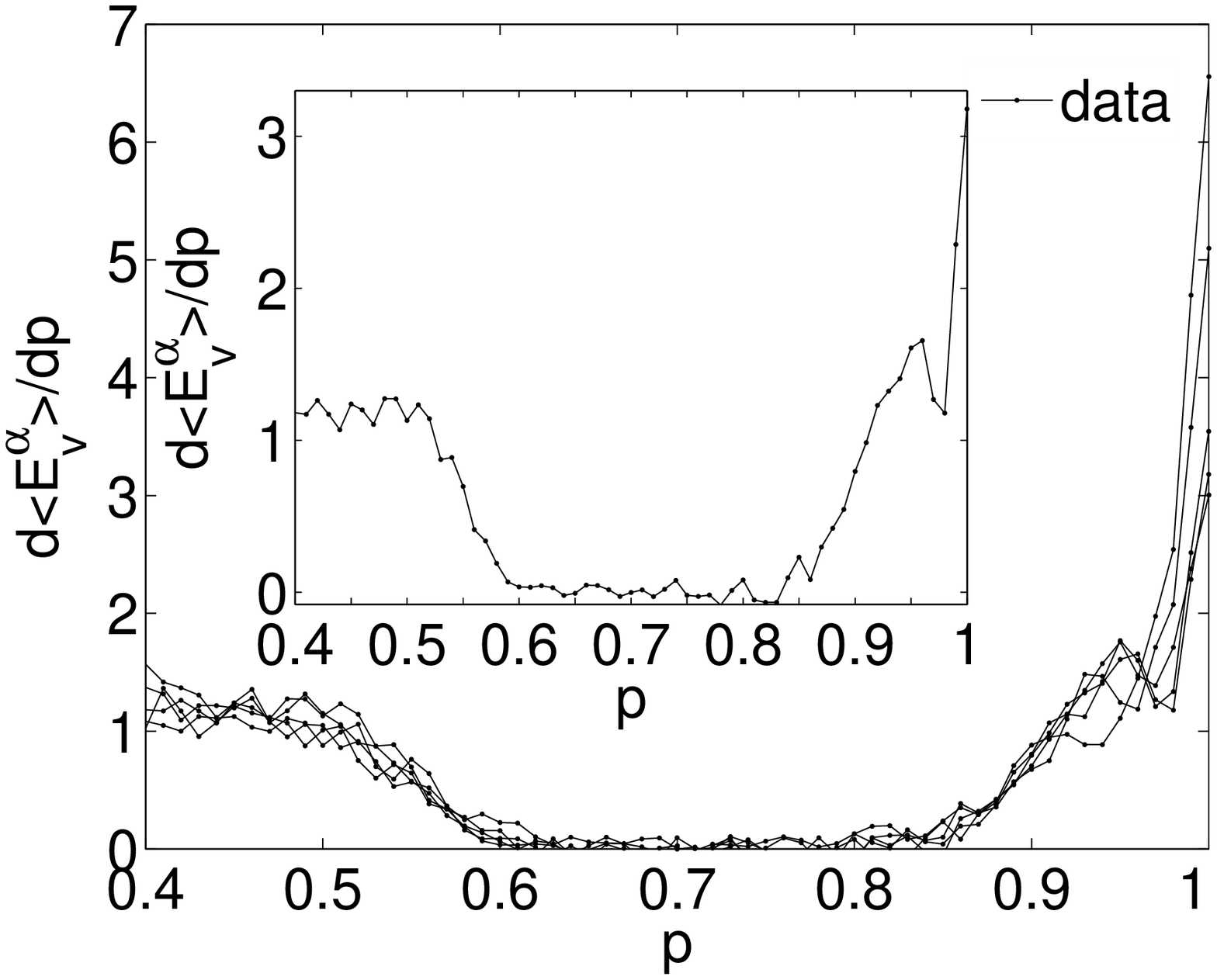}
 \caption{Some quantities for eigenstates with eigenenergies
 $\varepsilon_\alpha\in[0-0.02,0+0.02]$. (a) The
von Neumann entropy $\langle E^\alpha_v \rangle$  and (b) the
$d\langle E^\alpha_v \rangle/dp$ varying with  the site occupation
probability $p$. The system sizes $L=20,30,...,60$ and the arrow
direction in (a) and (b) denotes the increasing of $L$. The inset
in (b) is for $L=60$. }\label{Fig6}
\end{figure}

In the following,  we discuss the von Neumann entropy $\langle
E^\alpha_v \rangle$ for eigenstates with eigenenergies
 $\varepsilon_\alpha\in[0-0.02,0+0.02]$. The
von Neumann entropy $\langle E^\alpha_v \rangle$ and the
corresponding  derivative $d\langle E^\alpha_v \rangle/dp$ varying
with the site occupation probability $p$ are shown in
Figs.\ref{Fig6}(a) and (b), respectively. Fig.\ref{Fig6}(a) shows
that $\langle E^\alpha_v \rangle$ first increases with $p$ until
to a plateau at $p\in[0.6,0.85]$, then continues to increase,
which is quitely different from that shown in Fig.\ref{Fig5}(a)
for eigenenergies away from the band center. Fig.\ref{Fig6}(b)
shows that the derivative $d\langle E^\alpha_v \rangle/dp$
drastically decreases near $p\simeq0.6$ and drastically increases
near $p\simeq0.85$.

With a real-space renormalization-group method, Odagaki and Chang
observed three regimes of the electronic properties in QP models,
which divided by the classical percolation threshold $p_c$ and the
quantum percolation threshold $p_q$($p_q>p_c$) \cite{od84}. When
$p<p_c$, electrons cannot tunnel between different isolated
clusters and electrons are considered to be localized in the
classical sense even in the quantum case. When $p_c<p<p_q$, due to
quantum interference effects, electrons cannot spread infinitely
even there is an infinitely extended channel. The regime is called
quantum localization regimes. When $p>p_q$, electrons can spread
infinitely and electron states are extended. They found
$p_c=0.618$ and $p_q=0.867$. It is interesting that the von
Neumann entropy $\langle E^\alpha_v \rangle$ drastically changes
near the two $p$ values.



\subsection{\label{sec32}Phase diagram}

\begin{figure}
(a)\includegraphics[width=2.5in]{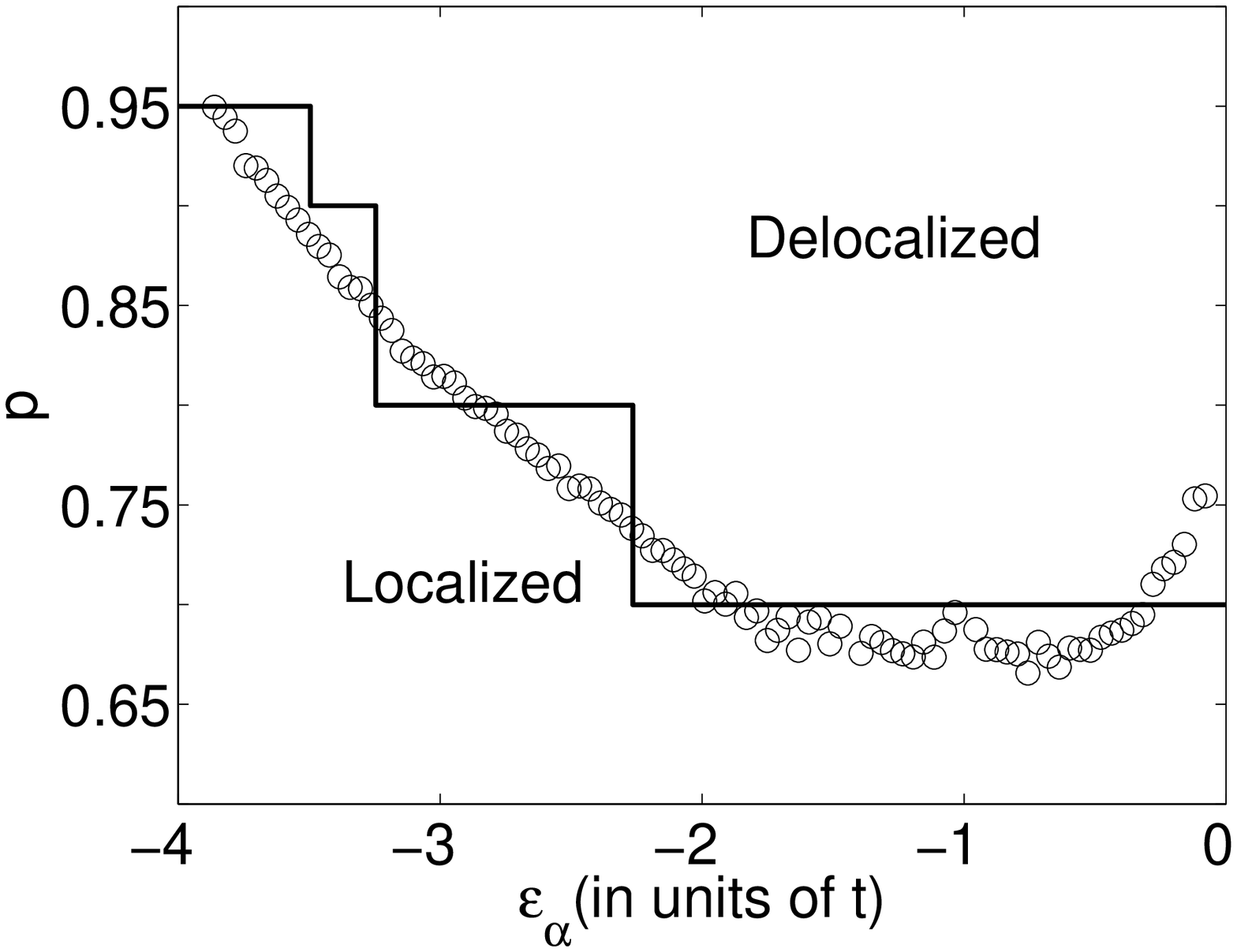}
(b)\includegraphics[width=2.5in]{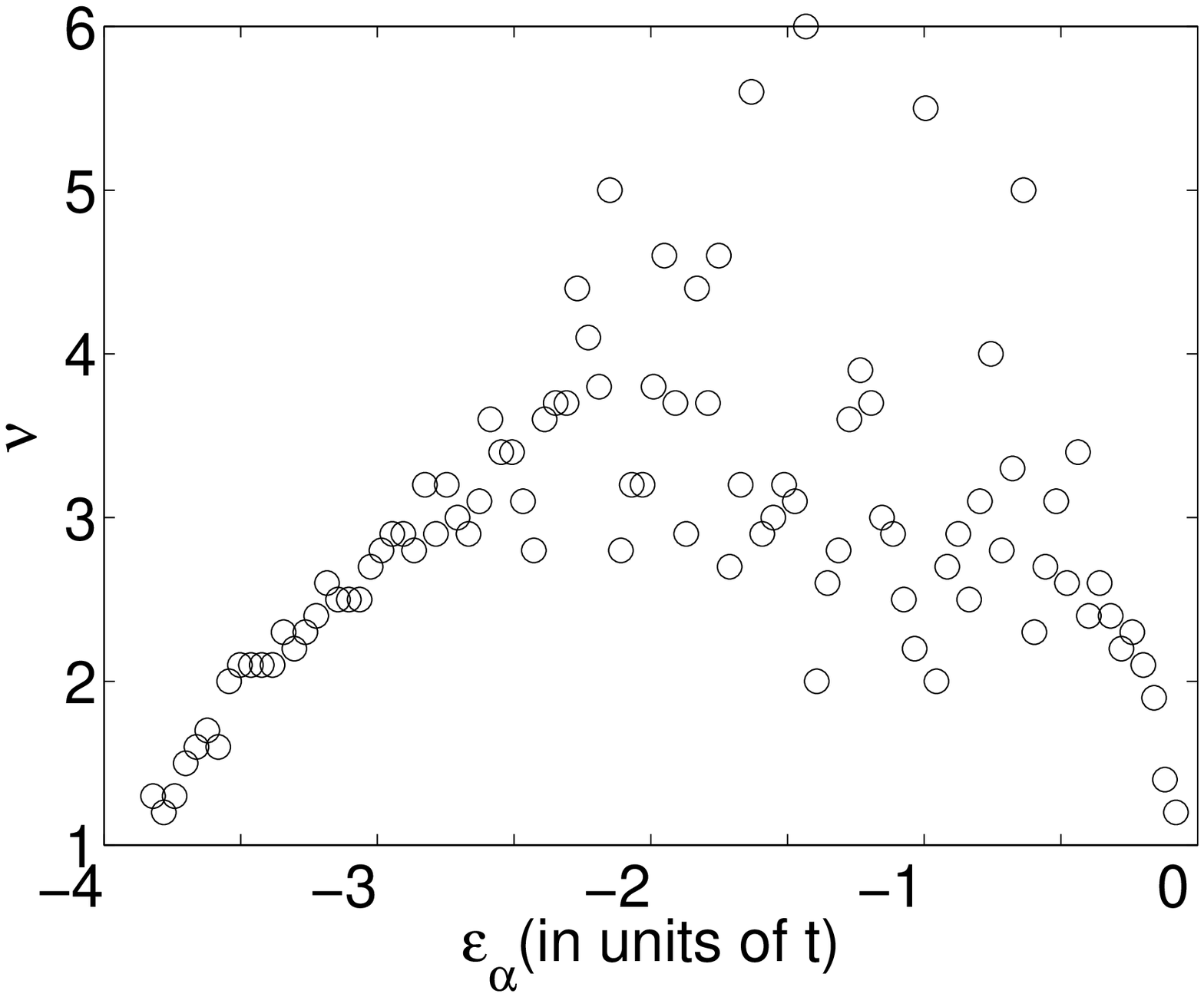}
(c)\includegraphics[width=2.5in]{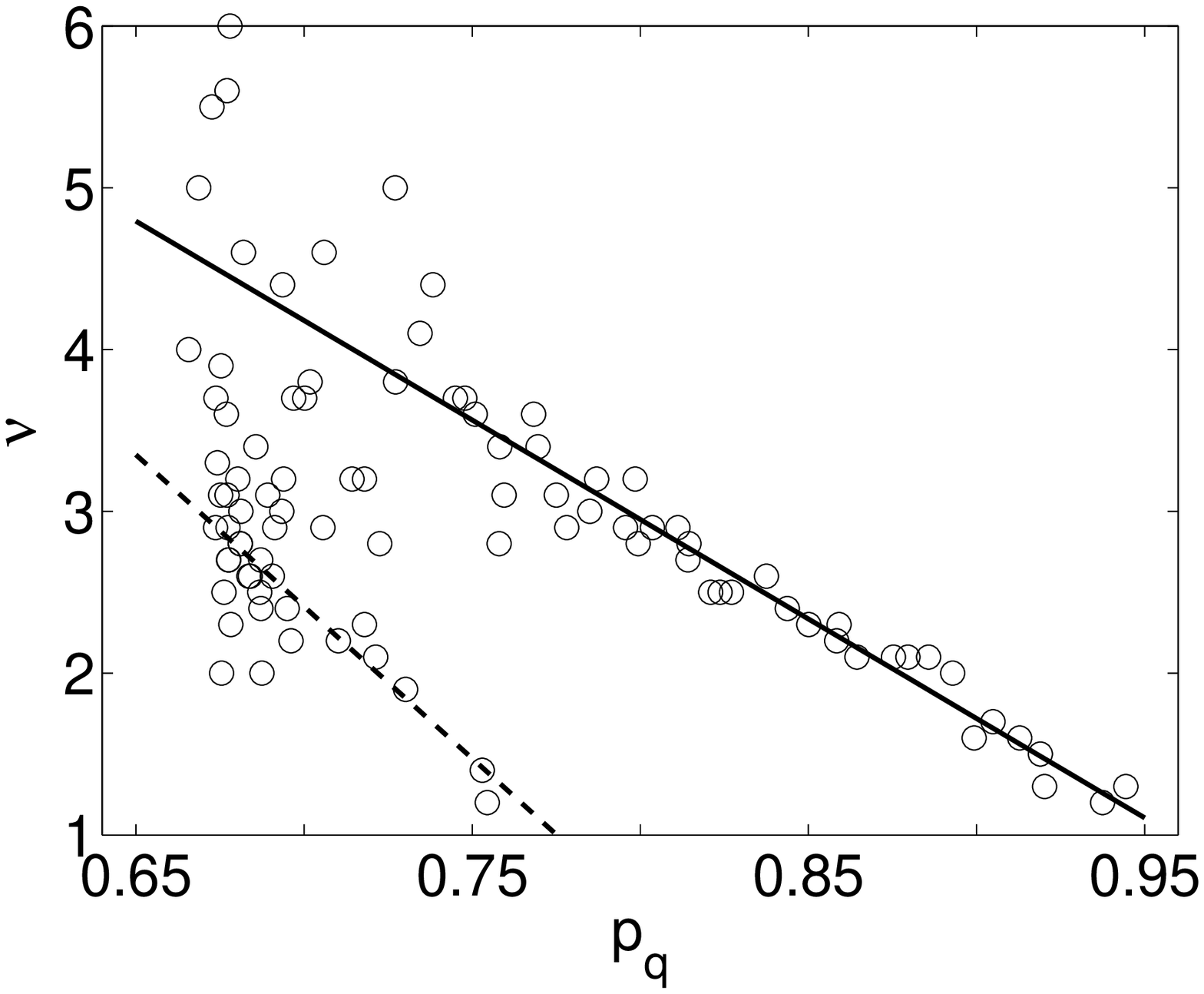} \caption{(a) The phase
diagram of LDTs in the space of eigenenergies $\varepsilon_\alpha$
and the site occupation probability $p$. The values of $p$ at the
LDTs are QP threshold  $p_q$. Line: mobility edge trajectory
obtained by Koslowski and von Niessen \cite{ko90}.  (b)The
critical exponents $\nu$ versus eigenenergies
$\varepsilon_\alpha$. (c)The critical exponents $\nu$ versus QP
threshold $p_q$. The solid line and the dashed line are the best
linear fitting for $\varepsilon_\alpha<-2.0t$ and
$\varepsilon_\alpha>-0.5t$, respectively. }\label{Fig7}
\end{figure}

We have studied and extensively analyzed the von Neumann entropy
for all the other eigenstates. As the system in Eq.(\ref{form3})
is bipartite, $\varepsilon_\alpha$ and $-\varepsilon_\alpha$ are
both eigenvalues of $\hat{H}_{AA}$ \cite{ko90}. Therefore we will
restrict our investigation to one (left) half side of the band,
i.e.,$-4<\varepsilon_\alpha<0$. Fig.\ref{Fig7}(a) presents the
phase diagram of LDTs in the $\varepsilon_\alpha-p$ plane.  The
values of $p$ at the LDTs are the QP threshold  $p_q$, which is
obtained by the extrapolation method similarly as that shown in
Fig.\ref{Fig5}(c). The trend for $p_q$ varying with
$\varepsilon_\alpha$ is similar as that for 2D quantum
bond-percolation models on square lattices \cite{is08} and 3D
quantum site-percolation models on simple cubic lattice
\cite{so87,sc05}. In detail, we observe the nonmonotonic
dependence of the values of $p_q$ on eigenenergies
$\varepsilon_\alpha$. Notice that in the region
$-2t<\varepsilon_\alpha<-0.5t$ the QP threshold $p_q$ are nearly
constant and a weak maximum at $\varepsilon_\alpha\simeq-t$, which
are very similar as that found in Ref.[\onlinecite{so87}]. That a
weak maximum for $p_q$ at $\varepsilon_\alpha\simeq-t$ has also
been found and discussed in Ref.[\onlinecite{sc05}], which may be
due to the existence of von Hove singularity at the energy
\cite{so87}.  We find all the QP threshold $p_q$ are greater than
the classical percolation threshold $p_c\simeq0.593$ and the
lowest value $p_q\simeq0.665$. As shown in Fig.\ref{Fig7}(a), the
phase diagram is very consistent with the mobility edge trajectory
shown in Ref.[\onlinecite{ko90}] for the same model obtained by
the Thouless-Edwards-Licciardello method.

Fig.\ref{Fig7}(b) presents all the critical exponents $\nu$ versus
eigenenergies $\varepsilon_\alpha$ according to the finite scaling
analysis similarly as that shown in Fig.\ref{Fig5}(d). We find the
values of $\nu$ depend on eigenenergies $\varepsilon_\alpha$. In
detail, in the region $-2t<\varepsilon_\alpha<-0.5t$ where $p_q$
are almost constant(see Fig.\ref{Fig7}(a)), most of the values of
$\nu$  are distributed in a relatively narrow interval
$[2.6,3.2]$. From the band edge to $\varepsilon_\alpha\simeq-2t$,
$\nu$ increase with $\varepsilon_\alpha$, while from
$\varepsilon_\alpha\simeq-0.5t$ to the band center, $\nu$ decrease
with $\varepsilon_\alpha$. All $\nu$ are larger than $2/D(D=2)$,
which satisfies the assumption that $\nu$ must satisfy the bound
$\nu\geq 2/D$ for random systems \cite{ch86}. Though $\nu$ has
been extensively studied in 3D QP models \cite{ko91,be96,od84}, to
our best knowledge, there are few works to study $\nu$ in 2D QP
models except in Ref. \onlinecite{od84}, where the critical
exponent for correlation lengths $\nu\simeq3.35$. The varying of
the critical exponent $\nu$ with the QP threshold $p_q$ is plotted
in Fig.\ref{Fig7}(c). It shows near the band edge and near the
band center the relation between $\nu$ and $p_q$ is linear,
respectively,  but the linear relations are different. This may be
caused by the different symmetry and/or degeneration at the two
energy regions \cite{ki72,sc05}.

\section{\label{sec4}Conclusions and Discussions}
In this paper, we have detailed studied the von Neumann entropy
$\langle E^\alpha_v \rangle$ varying with  eigenenergies
$\varepsilon_\alpha$ and accessible site concentrations $p$ by
quantum particles in two-dimensional quantum site-percolation
models on square lattices.

For Eigenenergies away from the band center,  we determine the QP
threshold $p_q$ by the derivative of von Neumann entropy is
maximal at the point. Based on this, we give the phase diagram of
LDTs in the $\varepsilon_\alpha-p$ plane, which is very consistent
with the mobility edge trajectory shown in
Ref.[\onlinecite{ko90}]. From the phase diagram, we observe the
non-monotonic eigenenergies dependence of $p_q$ and the lowest
value $p_q\simeq0.665$. At the same time, the finite-size scaling
analysis is performed at all LDTs points. To the best of our
knowledge, it is the first time to obtain all the critical
exponents $\nu$ at the whole energy space for 2D QP models. We
find the critical exponents $\nu$ depend on eigenenergies
$\varepsilon_\alpha$.

For eigenenergies near the band center, the variations of the von
Neumann entropy $\langle E^\alpha_v \rangle$  with respect to $p$
for are quitely different from that for eigenenergies away from
the band center. It can reflect the classical percolation
threshold $p_c$ and the QP threshold $p_q$ for eigenenergies at
the band center.

All our numerical results show that there is $p_q<1$ in 2D QP
models and $p_q$ depends on eigenenergies. The debates on the
values of $p_q$  may be partially due to different energies
treated in literatures.


\begin{acknowledgments}
This project was supported by the National Natural Science
Foundation of China (Grants No. 10674072, 10974097), by the
Specialized Research Fund for the Doctoral Program of Higher
Education (Grant No. 20060319007), and by National Key Projects
for Basic Research of China (Grant No. 2009CB929501). L.G. is
supported in part by the National Natural Science Foundation of
China (Grants No. 10904047), by the Nature Science Foundation of
Jiangsu Province of China (Grant No. 08KJB140005)and by the
Specialized Research Fund for the Doctoral Program of Higher
Education (Grant No. 20060293001).
\end{acknowledgments}

\end{document}